\documentclass[12pt]{iopart}

\newcommand{\perm}[0]{\mathrm{perm}}
\newcommand{\bra}[1]{\langle#1|}
\newcommand{\ket}[1]{|#1\rangle}

\usepackage{graphicx}
\usepackage{iopams}
\usepackage{color,soul}
\usepackage{comment}

\begin{document}

\title{Multiparameter estimation with single photons}
---Linearly-optically generated quantum entanglement beats the shotnoise limit

\author{Chenglong You$^1$, Sushovit Adhikari$^1$, Yuxi Chi$^2$, Margarite L. LaBorde$^1$, Corey T. Matyas$^3$, Chenyu Zhang$^4$, Zuen Su$^{5,6}$, Tim Byrnes$^{7,4,8}$, Chaoyang Lu$^{5,6}$, Jonathan P. Dowling$^{1,4}$ and Jonathan P. Olson$^9$}

\address{$^1$ Hearne Institute for Theoretical Physics, Department of Physics and Astronomy, Louisiana State University, Baton Rouge, Louisiana 70803, USA}
\address{$^2$ Institute of Solid Mechanics, Beihang University, Beijing 100191, China}
\address{$^3$School of Electrical Engineering and Computer Science, Louisiana State University, Baton Rouge, Louisiana 70803 }
\address{$^4$ New York University Shanghai, 1555 Century Ave, Shanghai 200122, China}
\address{$^5$ Hefei National Laboratory for Physical Sciences at Microscale and Department of Modern Physics, University of Science and Technology of China, Hefei, Anhui 230026, China}
\address{$^6$ CAS Centre for Excellence and Synergetic Innovation Centre in Quantum Information and Quantum Physics, University of Science and Technology of China, Hefei, Anhui 230026, China}
\address{$^7$ State Key Laboratory of Precision Spectroscopy, School of Physical and Material Sciences, East China Normal University, Shanghai 200062, China}
\address{$^8$ NYU-ECNU Institute of Physics at NYU Shanghai,
3663 Zhongshan Road North, Shanghai 200062, China}
\address{$^9$ Department of Chemistry and Chemical Biology, Harvard University, Cambridge, Massachusetts 02138, USA}

\ead{olson.jonathanp@gmail.com}
\vspace{10pt}
\begin{indented}
\item[]\today
\end{indented}

\begin{abstract}
It was suggested in Ref. [Phys. Rev. Lett. 114, 170802] that optical networks with relatively inexpensive overhead---single photon Fock states, passive optical elements, and single photon detection---can show significant improvements over classical strategies for single-parameter estimation, when the number of modes in the network is small ($n<7$).  A similar case was made in Ref. [Phys. Rev. Lett. 111, 070403] for multi-parameter estimation, where measurement is instead made using photon-number resolving detectors.  In this paper, we analytically compute the quantum Cram\'er-Rao bound to show these networks can have a constant-factor quantum advantage in multi-parameter estimation for even large number of modes.  Additionally, we provide a simplified measurement scheme using only single-photon (on-off) detectors that is capable of approximately obtaining this sensitivity for a small number of modes.
\end{abstract}

%
\vspace{2pc}

%
\submitto{\JOPT}
%
%
%

\section{Introduction}

Phase parameter estimation with optical interferometry has long been a cornerstone for studying systems of both theoretical and practical interest, even as early as the Michelson-Morley experiment in 1887.  Since the discovery of quantum optical interferometry, it has been shown that strategies utilizing non-classical states of light can be used for both single and multiple parameter estimation to theoretically allow for advantages in precision over classical methods \cite{Humphreys13,Lee02,Ciampini16,Dowling08, Giovannetti11, Lloyd06, Caves08,Gard2017}.  These quantum advantages are of particular interest for applications where the target is especially photosensitive, such as in the imaging of biological tissue, where estimating multiple parameters simultaneously is a fundamental task \cite{Warwick13,Sandoghdar14,Andrews84, Sandoghdar11}. Unfortunately, many of the quantum states required to enable these strategies are notoriously difficult to create or are extremely sensitive to noise, therefore limiting the systems of interest in which quantum optical interferometry might be realistically advantageous \cite{Dowling07RMP, Dowling07PRL, DEM15, Dem12, Walmsley09, gard2016photon,Anisimov10}.  The search continues to find architectures and states, which can achieve some level of super-sensitivity (beating the equivalent of the shot-noise limit), but can be readily made and are robust to noise.

Substantial progress has recently been made in the development of on-demand single-photon sources. 
Among other applications of these sources, experimental
BosonSampling is claimed to be a leading candidate for showing post-classical computation \cite{AaronsonBS,gard2015introduction,Spring798,Broome794,Crespi13,Spagnolo14,Lund17,Pan17}.  This has been a major motivating factor for a renewed interest in linear optical systems, although the claims for demonstrating imminent quantum supremacy is still under debate \cite{neville2017no,Clifford17}. These sources, together with high-efficiency detectors and waveguides, which can be integrated onto an all-optical chip, allow for an impressive level of fidelity in comparison to networks utilizing nonlinear optical elements and photon-number resolving detectors that are often necessary for implementing quantum metrology architectures. It was recently shown in Refs.~\cite{olson16,Motes15} that multimode interferometric devices comprised of only these simple linear components, including quantum Fourier transform interferometers (QUFTI), can be used to achieve super-sensitivity for single parameter estimation.  A similar device with photon-number resolving measurements was shown to be supersensitive for multiparameter estimation \cite{Humphreys13}.  In each case, however, the maximum number of modes which admitted an improvement over the shot-noise limit was small.

In this paper, we consider an analogous architecture that admits super-sensitivity for multi-parameter estimation while maintaining a relatively modest experimental overhead. Our analytic computation of the Fisher information for our device (Fig. \ref{fig:arch}) shows that the sensitivity continues to beat the shot-noise limit---even in the limit of a large number of modes.  Additionally, we show that our designs scale surprisingly well under photon loss and non-deterministic photon production from a source.

\section{Multi-parameter Estimation in a Parallel QuFTI}
\begin{figure}
\begin{center}
\includegraphics[width=0.6\columnwidth]{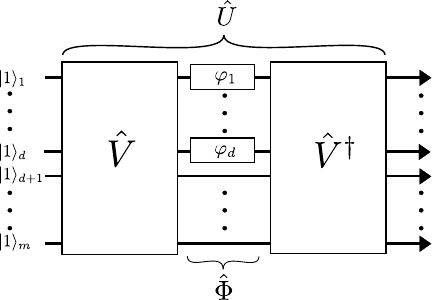} %
\caption{\label{fig:arch}Architecture of the proposed parallel QuFTI optical interferometer, which simultaneously measures $d$ independent unknown phases $\{\varphi_j\}_{j=1}^d$.  The interferometer consists of $m$ modes with an input of $m$ single photons, $\ket{1}^{\otimes m}$.  The unitary $\hat{V}$ (and its conjugate) is a quantum Fourier transform implemented with a network of beamsplitters and phase shifters.  Several detection strategies for the output are presented in Section 3.}
\end{center}
\end{figure}

In this current work we consider an architecture for an interferometer similar to our single parameter estimation strategy originally described in Ref.~\cite{Motes15}, 
where instead we now consider an estimate of multiple independent phases simultaneously. The interferometer consists of $m$ mode with a photon in each mode with input $\left| {\left. {{\psi _{\rm{in}}}} \right\rangle  = } \right.{\left| {\left. 1 \right\rangle } \right.^{ \otimes m}}$, as shown in Fig.~\ref{fig:arch}.  The input is fed into a particular passive linear optical unitary $\hat{U}= 
\hat{V}\hat{\Phi} \hat{V}^{\dagger}$, where $\hat{V}=\{V_{ij}\}$ is the quantum Fourier transform
\begin{equation}
V_{ij}=\frac{1}{\sqrt{m}}e^{2\pi(i-1)(j-1)/m},
\end{equation}
and $\hat{\Phi}=\{\Phi_{k\ell}\}$ is a $m\times m $ diagonal matrix of $d$ independent phases $\vec{\boldsymbol{\varphi}}=\{\varphi_j\}_{j=1}^d$ which we would like to estimate.  $\hat{\Phi}$ is diagonal and has the form,
\begin{equation}
\Phi_{k\ell}=
\begin{array}{lc}
  \Bigg\{ & 
    \begin{array}{lc}
      \delta_{k\ell}\cdot e^{i\varphi_k} & k\leq d \\ \delta_{k\ell} & k>d
    \end{array}
\end{array}.
\end{equation}
Other than the form of $\hat{\Phi}$, the above is identical to our previous QuFTI of Ref.~\cite{Motes15}, which leads us to refer to this device as a ``parallel QuFTI". In section 3, we will consider several different measurement strategies ranging from photon counting with number-resolution to on-off photodetection, which only distinguishes vacuum from a non-zero number of photons.  For each strategy, the resulting probability distribution obtained from repeated measurements then acts as a measure of the unknown phases.

The output state $\ket{\psi_{\rm{out}}}$ of the interferometer is, 
\begin{equation}
	\ket{{ \psi  }_{ \rm{out} }}=\hat{U}\ket{{ \psi  }_{\rm{in}}}=\sum _{ i }^{  }{ { \gamma  }^{(i) } } \ket{{ n }_{ 1 }^{ (i) },\dots ,{ n }_{ m }^{ (i) }}=\sum _{ i }^{  }{ { \gamma  }^{ (i) } }\ket{\boldsymbol{n}^{(i)}},
\end{equation}
where the sum is over all possible output photon configurations $\ket{\boldsymbol{n}^{(i)}}=\ket{{ n }_{ 1 }^{ (i) },\dots ,{ n }_{ m }^{ (i) }}$ with $m$ total photons, i.e. $\sum_{j}n_j^{(i)}=m$.

The coefficients ${ \gamma }^{ (i) }$ of every output configuration are related to matrix permanents of matrices closely related to $\hat{U}$ \cite{olsonthesis16}.  More precisely, for the photon configuration $i$ and associated matrix permanent $\perm(W^{(i)})$, if we denote the $j$th row vector of $\hat{U}$ as $\textbf{u}_j$, then $W^{(i)}$ consists of $n^{(i)}_j$ rows of $\textbf{u}_j$ (note that matrix permanents are invariant under row interchange, so the ordering of the rows is unimportant).  The corresponding coefficient is given by,
\begin{equation}
	{ \gamma }^{ (i) }=\frac { \perm({ W }^{ (i) }) }{ \sqrt { { n }_{ 1 }^{ (i) }!\dots { n }_{ m }^{ (i) }! } }.
    \label{eq:gamma}
\end{equation}
Note that although the computational complexity of matrix permanents is in general $\# P$-hard to compute, even in the average case, matrices with certain symmetries may still be tractable, such as the configuration $\ket{\boldsymbol{n}}=\ket{1,1,\dots, 1}$ when $d=1$ \cite{olson16}.

Recall that our goal is to use the interferometer described above to estimate the $d$ unknown phases $\vec{\boldsymbol{\varphi}}$.  For a given measurement scheme, the Cram\'er-Rao bound limits the precision of an estimate via the inequality,
\begin{equation}
	{ \left| \Delta \vec{\boldsymbol{\varphi}} \right| }^{ 2 }\equiv \sum _{ j=1 }^{ d }{ \Delta { \varphi }_{ j }^{ 2 } } \equiv \textup{Tr}[\textup{Cov}(\vec{\boldsymbol{\varphi}} )]\ge \frac { 1 }{ \nu  }\textup{Tr}[{ { \mathcal{F} } } _{\vec{\boldsymbol{\varphi}}}^{ -1 }] \label{eq:crb}
\end{equation}
where $\nu$ is the number of independent trials, and the matrix ${ { \mathcal{F} } } _{\vec{\boldsymbol{\varphi}}}=\{\mathcal{F}_{i,j}^{\rm{clas}}\}$ is the classical Fisher information given by,
\begin{equation}
\mathcal{F}_{ i,j }^{\rm{clas}}=\sum _{ x }^{ }{ \frac { 1 }{ p(x|\vec{\boldsymbol{\varphi}} ) } \frac { \partial p(x|\vec{\boldsymbol{\varphi}} ) }{ \partial { \varphi }_{ i } } \frac { \partial p(x|\vec{\boldsymbol{\varphi}} ) }{ \partial { \varphi }_{ j } } },
\end{equation}
where $p(x|\vec{\boldsymbol{\varphi}})$ is the probability of observing outcome $x$ conditioned on $\vec{\boldsymbol{\varphi}}$.  Because of the dependence of the Fisher information on $\vec{\boldsymbol{\varphi}}$, it may be the case that the measurement precision is best near certain values of $\vec{\boldsymbol{\varphi}}$, as in Refs. \cite{olson16,Bentivegnae15}.

The quantum Cram\'er-Rao bound (QCRB)\cite{meystre1977quantum} lower-bounds the uncertainty of estimating   $\vec{\boldsymbol{\varphi}}$ from a given \textit{quantum} state that encodes information about $\vec{\boldsymbol{\varphi}}$, but is independent of any measurement scheme and dependent only on the probe state
.  The QCRB is identical to Eq.~(\ref{eq:crb}), except that the Fisher information matrix $\mathcal{F}_{ i,j }^{\rm{clas}}$ is replaced by the quantum Fisher information (QFI) matrix \cite{meystre1977quantum},
\begin{equation}
	\mathcal{F}_{ i,j }^{\rm{quant}}=\frac{1}{2} \bra{{ \psi  }_{ \rm{out} }}(L_i L_j+ L_j L_i)\ket{{ \psi  }_{ \rm{out} }}.
\end{equation}
where $L_i=2(\ket{\partial_{{ \varphi }_{ i }}{ \psi  }_{\rm{out} }}\bra{{ \psi  }_{ \rm{out} }}+\ket{{ \psi  }_{ \rm{out} }}\bra{\partial_{{ \varphi }_{ i }}{ \psi  }_{ \rm{out} }})$.  Subsequently, we will refer to $\mathcal{F}_{\vec{\boldsymbol{\varphi}}}=\{\mathcal{F}_{ i,j }^{\rm{quant}}\}$ to mean the QFI matrix. It is worth noting that the dimension of both the Fisher information matrix and the QFI matrix is equal to the number of phases we are estimating ($d$ in our case).

It was shown by Humphreys et al.~\cite{Humphreys13} that for arbitrary pure input states of multi-mode Fock states, the QFI matrix of the estimated phases is given as
\begin{equation}
	\mathcal{F}_{\vec{\boldsymbol{\varphi}}}=4\sum _{ i }^{ }{ { \left| { \gamma }^{ (i) } \right| }^{ 2 } } \ket{{ \boldsymbol{n} }^{ (i) }}\bra{{ \boldsymbol{n} }^{ (i) }}-4\sum _{ i,j }^{ }{ { \left| { \gamma }^{ (i) } \right| }^{ 2 }{ \left| { \gamma }^{ (j) } \right| }^{ 2 } } \ket{{ \boldsymbol{n} }^{ (i) }}\bra{{ \boldsymbol{n} }^{ (j)}},
\end{equation}
where the $\gamma^{(i)}$ are defined in Eq.~(\ref{eq:gamma}).
The quantum Fisher Information matrix can be calculated as,

\begin{equation}
	[\mathcal{F}_{\vec{\boldsymbol{\varphi}}}]_{l,n}=4\left< \hat b_l^\dag {\hat b_l}\hat b_n^\dag {\hat b_n} \right> -4 \left< \hat b_l^\dag {\hat b_l} \right>\left< \hat b_n^\dag {\hat b_n} \right>,
\end{equation}
where $\hat b_i^{\dagger} = \sum_j V_{i,j} \hat a_{j}^{\dagger} $ \cite{vogel2006quantum}.

Calculating the QFI for the setup with a $k$-photon Fock state in every mode, we obtain,
\begin{equation}
\mathcal{F}_{\vec{\boldsymbol{\varphi}}}=4k(k + 1)\left( {\begin{array}{*{20}{c}}
{\frac{{m - 1}}{m}}&{ - \frac{1}{m}}& \cdots &{ - \frac{1}{m}}\\
{ - \frac{1}{m}}&{\frac{{m - 1}}{m}}& \cdots &{ - \frac{1}{m}}\\
 \vdots & \vdots & \ddots & \vdots \\
{ - \frac{1}{m}}&{ - \frac{1}{m}}& \cdots &{\frac{{m - 1}}{m}}
\end{array}} \right),
\end{equation}
Details of the calculation can be found in Appendix A. Computing its inverse \cite{inversematrix}, we find a similarly patterned matrix, 

\begin{equation}
\mathcal{F}_{\vec{\boldsymbol{\varphi}}}^{-1}=\frac{1}{{4k(k + 1)}}\left( {\begin{array}{*{20}{c}}
{\frac{{m - d + 1}}{{m - d}}}&{\frac{1}{{m - d}}}& \cdots &{\frac{1}{{m - d}}}\\
{\frac{1}{m-d}}&{\frac{{m - d + 1}}{{m - d}}}& \cdots &{\frac{1}{{m - d}}}\\
 \vdots & \vdots & \ddots & \vdots \\
{\frac{1}{{m - d}}}&{\frac{1}{{m - d}}}& \cdots &{\frac{{m - d + 1}}{{m - d}}}
\end{array}} \right).
\end{equation}
Substituting the trace of $\mathcal{F}_{\vec{\boldsymbol{\varphi}}}^{-1}$ into Eq.~(\ref{eq:crb}) and recalling the matrix is $d\times d$, we arrive at the bound,
\begin{equation}
{ \left| \Delta \vec{\boldsymbol{\varphi}} \right| }^{ 2 }\geq\frac { 1 }{ {\nu} } \frac{1}{4k(k+1)}\frac { d(m-d+1) }{ (m-d) }. \label{eq:fisher}
\end{equation}
In this paper, we will consider the case that $k=1$, since on-demand single-photon sources are quickly becoming experimentally viable. While larger Fock state generation remains a challenge \cite{motherfocker}, it is interesting to see that, since $\left| \Delta \vec{\boldsymbol{\varphi}} \right|^2$ scales inversely with $k^2$, indicating that an asymptotic improvement approaching the Heisenberg limit is possible if such states could be easily prepared. 

\section{Measurement Strategies}
To examine the sensitivity of the described multi-arm interferometer, we will first compare the QFI for several different phase estimation strategies, assuming that the QCRB can be saturated in each case.  Suppose there are $m$ modes, and we wish to simultaneously measure $d$ phases, where $d<m$ (at least one arm must be used as a reference).  For $k=1$, Eq.~ (\ref{eq:fisher}) reduces to, 
\begin{equation}\label{BS3totalvar}
{ \left| \Delta \vec{\boldsymbol{\varphi}}_1 \right| }^{ 2 }=\frac { 1 }{ {\nu}_{1} } \frac { (m-d+1)d }{ 8(m-d) },
\end{equation}
where $\nu_1$ denotes the number of measurements for the parallel QuFTI. 

To show any advantage for our scheme, we must compare our setup to other relevant architectures, which are limited in the same resources. We may consider a strategy using an identical interferometer, except where only a single phase is estimated at a time.  Such a strategy is simply a sequential version of the scheme developed by Olson et al. \cite{olson16}. We will refer this comparator scheme as ``sequential QUMI" for the remainder of this paper. Another comparison is made to the classical strategy where the inputs are uncorrelated coherent states $\otimes _{i = 1}^m\left| {{\alpha _i}} \right\rangle $. 

To make a fair comparison, we restrict that these three different schemes should use the same amount of photons. For the sequential QUMI, since we have $d$ phases to measure, and we need at least $m$ photons for a measurement of each phase, the variance becomes, 
\begin{equation}
	{ \left| \Delta \vec{\boldsymbol{\varphi}}_2 \right| }^{ 2 }=\frac { 1 }{ {\nu}_{2} }{ \left( \frac { 1 }{ \sqrt { 8\left( 1-\frac { 1 }{ m } \right) } } \right) }^{ 2 }d=\frac { 1 }{ {\nu}_{2} } \frac { m d }{ 8\left( m-1 \right) },
    \label{phi2}
\end{equation}
where $\nu_2$ is the number of repetitions of this protocol, and we have used the result of Ref.~\cite{olson16} to compute the sensitivity.  For $\nu_2=1$, the total number of photons used in sequential QUMI measurement is $m d$.

For the parallel QuFTI, a single measurement requires $m$ photons. Thus, for a fair comparison against the sequential QUMI, we require $\nu_1=d\nu_2$ so that Eq.~({\ref{BS3totalvar}}) becomes,
\begin{equation}\label{BS3totalvar1}
{ \left| \Delta \vec{\boldsymbol{\varphi}}_1 \right| }^{ 2 }=\frac { 1 }{ {\nu}_{1} } \frac { (m-d+1)d }{ 8(m-d) }
=\frac{1}{\nu_2}\frac { (m-d+1) }{ 8(m-d) }.
\label{phi1}
\end{equation}
Finally, for the classical strategy, we let the average photon number of the input $\bar{n}=\sum_{i=1}^m |\alpha_i|^2=md$ so that a fair comparison requires $\nu_{3}=\nu_{2}$, and the variance is \cite{Humphreys13}, 
\begin{equation}
	{ \left| \Delta \vec{\boldsymbol{\varphi}}_3 \right| }^{ 2 }=\frac{1}{\nu_{3}}\frac { { d }^{ 2 } }{ md }=\frac{1}{\nu_2}\frac{d}{m}.
    \label{phi3}
\end{equation}

\begin{figure}[ht]
\begin{center}
\includegraphics[width=\columnwidth]{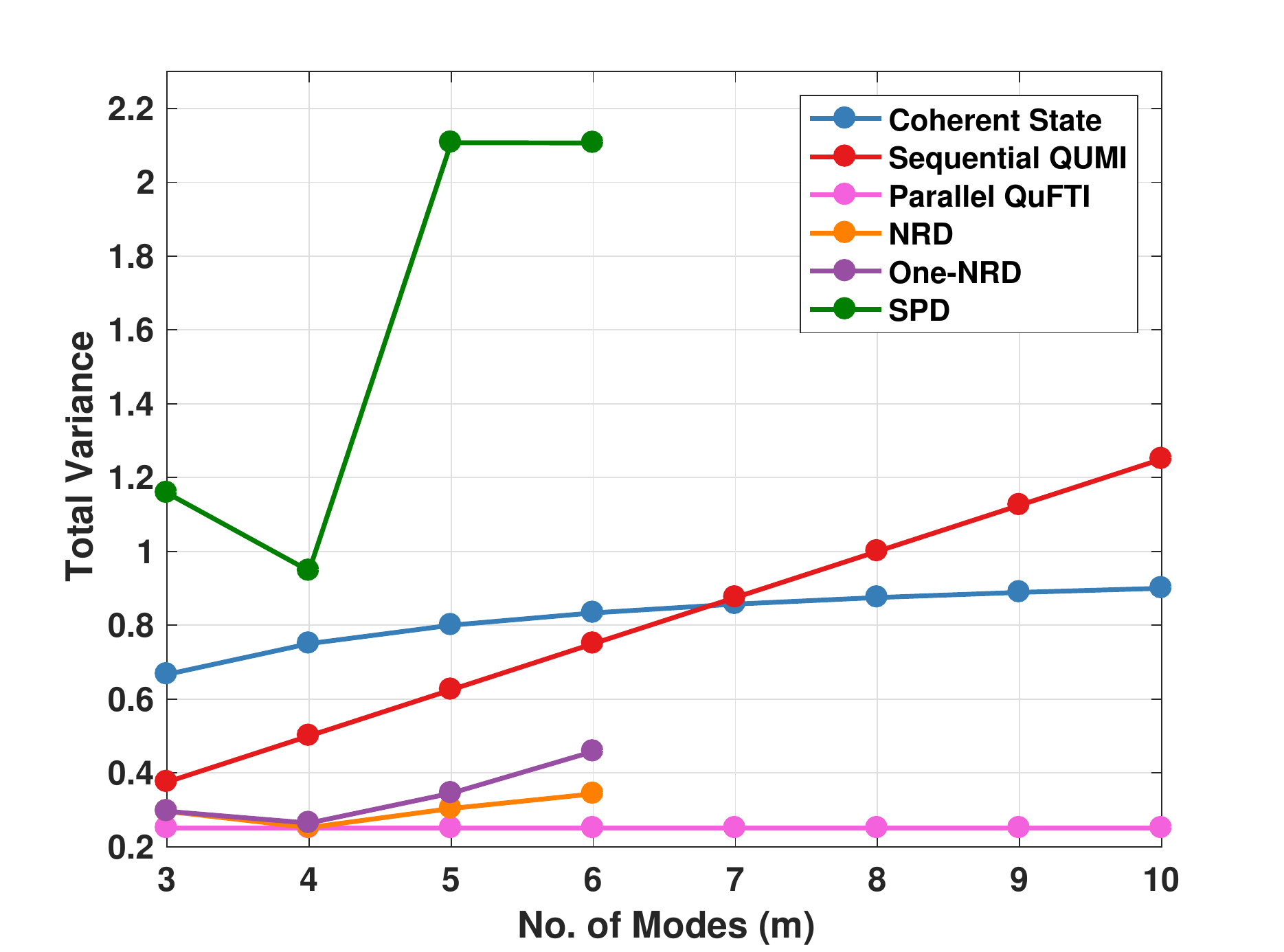} %
\caption{\label{fig:mordorplot} Total variance ($\Delta{\boldsymbol{\vec{\varphi}}^{2}}$) with different meteorological strategies when estimating multiple ($d=m-1$) parameters. The QCRB for the parallel QuFTI strategy (pink, Eq.~\ref{phi1}) gives the lowerbound on the variance for any measurement scheme. The One-NRD (purple), SPD (green) and NRD (orange) are obtained from numerically optimizing $\boldsymbol{\varphi}$ from the classical CRB (Eq.~\ref{eq:crb}). For comparison, the coherent state strategy (blue, Eq.~\ref{phi3}) and sequential QUMI (red, Eq.~\ref{phi2}) are shown.}

\end{center}
\end{figure}

%

Now that the variance of each strategy is expressed in terms of the same number of photons (namely, $md\nu_2$), we can easily compare them.  In the case that $d=1$, the sequential QUMI and parallel QuFTI are identical strategies, and the comparison against the classical strategy mirrors the prior result from Ref. \cite{olson16}, which showed an improvement over the classical strategy only for $m<7$.  However, as we scale up $d$ along with $m$, we see that the parallel QuFTI, the scheme we propose in this paper, continues to improve relative to the classical strategy.  Indeed, setting $d=m-1$ yields the maximum improvement over the classical case, where our parallel QuFTI achieves an asymptotic improvement of a factor of four in the total variance (see Fig.~\ref{fig:mordorplot}).

However, one should provide an actual detection scheme, rather than only providing a QFI calculation alone, to make a useful comparison.  This is because, while the QFI gives a theoretically attainable bound on the sensitivity, we wish to consider detection schemes, which can be realistically implemented \cite{Masahiro}.  For the sequential QUMI and coherent state strategies, single-photon detectors (SPD) and homodyne detection make up the QCRB-saturating measurement schemes, respectively.  For the parallel QuFTI, we will consider several cases.  Note that to compute the sensitivity of these specific detection schemes, we numerically compute the minimum of the classical Fisher information $\mathcal{F}^{\rm{clas}}$ corresponding to these schemes. However, numerically computing these values for a large number of modes was problematic due to the complex landscape optimization of the Fisher information. In addition to the overhead of calculating the matrix permanents, the optimization showed a sensitive dependence to the phases making it a numerically intensive task. 

First, we see that a detection scheme corresponding to an array of $m$ photon number-resolving detectors (NRDs) nearly achieves the QCRB of our parallel QuFTI (for the values we were able to compute).  However, NRDs are known to be far more costly and difficult to implement experimentally than SPDs. Unfortunately, while an array of SPDs performs well for the QUMI, they perform quite poorly for estimating multiple phases simultaneously, even for small $m$.  

To bridge the gap between these two measurement schemes, we propose a new measurement scheme which is far less experimentally demanding. We consider a combination of a single NRD in one arm together with SPDs in the remaining arms. Using this scheme, we note a sensitivity at par with the NRD case for small numbers of modes. One can see why this may be the case for a small number of modes---because of the symmetry of the QFT, regardless of the phases, any cyclic permutation of event outcomes are equally likely (for instance, if $m=3$, the (1,2,0), (0,1,2) and (2,0,1) outcomes occur with the same frequency). Of course, with the increase in the number of modes, the number of distinguishable events reduces, and we expect the sensitivity to worsen if we do not include more NRDs.  Furthermore, the presence of a single NRD can be approximated experimentally by mixing the target mode with a series of vacuum modes using beamsplitters, and placing SPDs at the output of each of these modes, as was done in Ref. \cite{Dovrat}.  

\section{Probabilistic Photon Sources}

The parallel QuFTI, particularly for a small number of modes with few NRDs, is readily implementable in a laboratory with available technology. One of the main requirements needed for our scheme is the generation of indistinguishable photons. There have been many proposals for single photon sources using atoms \cite{Rempe07}, molecules \cite{Sandoghdar10}, color centers in diamond \cite{Weinfurter00}, quantum dots \cite{Pan17,Shields07}, and spontaneous parametric down conversion (SPDC) \cite{Bentivegnae15}. Because many of these techniques produce single photons probabilistically, an input state consisting of $m$ photons is not always guaranteed.
\begin{figure}[ht]
\begin{center}
\includegraphics[width=\columnwidth]{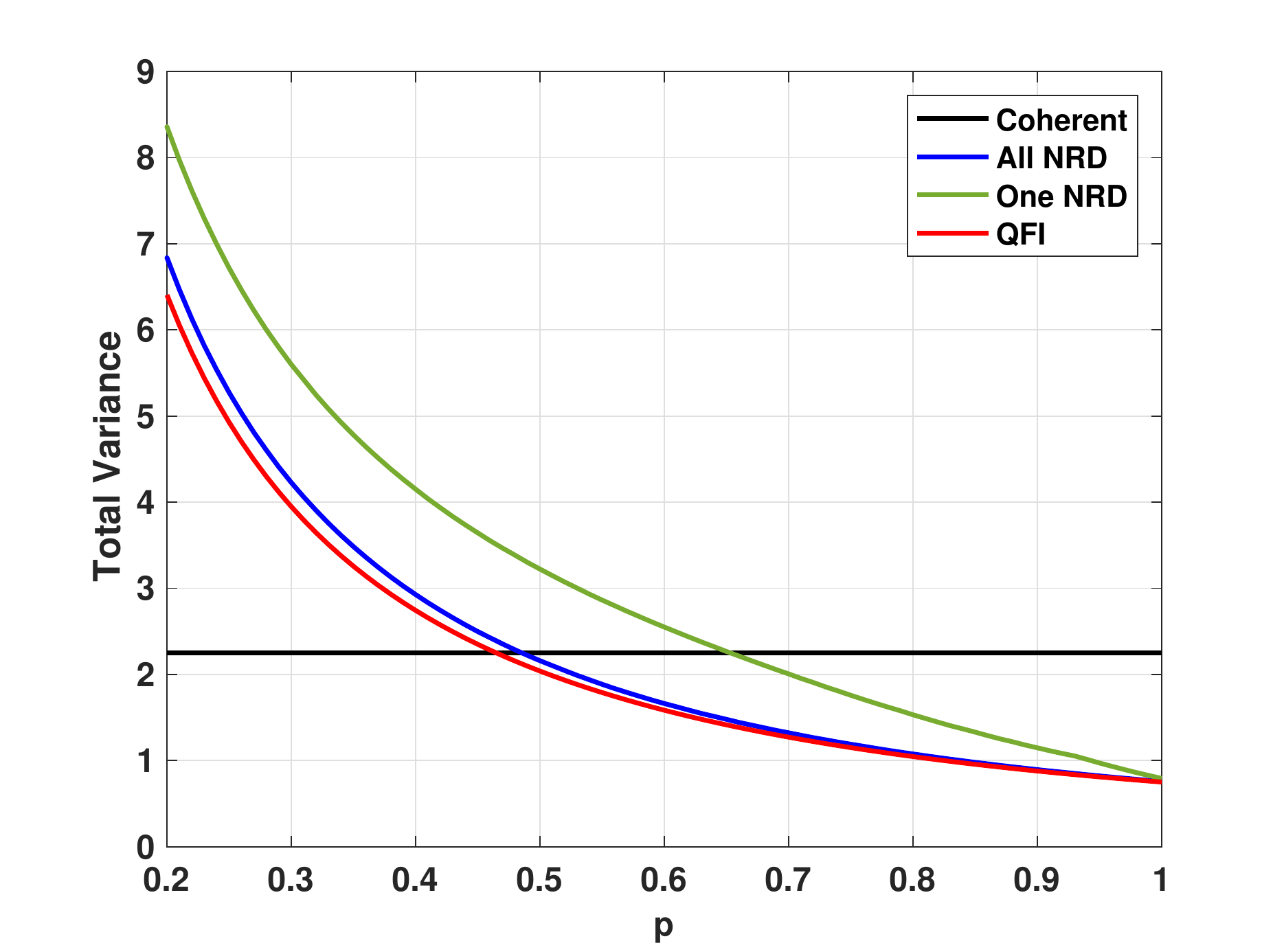} %
\caption{\label{fig:Scattershot}Total variance ($\Delta{\boldsymbol{\vec{\varphi}}_{\rm{avg}}^{2}}$) for scattershot four-mode, three-phase parallel QuFTI using all NRD detection scheme (blue line) and one-NRD detection scheme (green line) with photon source efficiency $p$ compared to the minimum variance for a lossless coherent source (black) with average photon number $\bar{n}=4$.}
\end{center}
\end{figure} 
Although it may be expected that truly on-demand sources will be available in the near future, we nonetheless consider a ``scattershot" input state to take into account the probabilistic nature of photon generation. A similar approach was recently proposed and demonstrated to improve the sampling efficiency for \textsc{BosonSampling} \cite{Bentivegnae15, Ralph14}. In an analogous way, we show that our scheme can still provide a sub-shot-noise sensitivity even when the photon sources are not necessarily reliable on-demand sources.  

In a scattershot scenario, photon pairs are emitted from a source (for instance, a SPDC) with some non-unit probability. A detection event of the one photon heralds the injection of the twin photon into a specific port of the interferometer. In this way, at a given time, one can keep track of the modes which received an input photon and the total number of photons present inside the interferometer.  With knowledge of the input, one can still make inferences about the phase, albeit with a lower sensitivity than an input with a full array of $m$ deterministic photon sources.

Consider a source of $m$ SPDC where the probability of generating a particular input configuration is $p_{i}$. For each input configuration, one can compute the associated variance $\Delta{\boldsymbol{\vec{\varphi}}_{i}^{2}}$ from the classical Fisher information, so that the average variance $\Delta{\boldsymbol{\vec{\varphi}}_{\rm{avg}}^{2}}$ is given by,
\begin{equation}
\Delta{\boldsymbol{\vec{\varphi}}} _{\rm{avg}}^{ - 2} = \sum\limits_{i = 1} {{p_i}} \Delta{\boldsymbol{\vec{\varphi}}_i^{ - 2}},
\end{equation}
where the summation is over the total number of input configurations.

We numerically consider the case of a four-mode, three-phase parallel QuFTI with probabilistic photon sources, where for simplicity all sources have an equal probability $p$ of emitting a heralded photon. As one can see in Fig.~\ref{fig:Scattershot}, even for a source efficiency around $50\%$, one can still beat a lossless coherent source, if one possesses a full NRD measurement. Yet even for a single NRD, a source of $65\%$ efficiency can still achieve supersensitivity. 

\section{Conclusion}

In this paper, we have considered a passive multi-mode interferometer for multiparameter phase estimation. We have shown that the quantum Cram\'er-Rao bound admits an asymptotic constant factor improvement in the sensitivity by a factor of 4, which can be approximately obtained for a small number of modes with an array of single photon detectors and only one number-resolving detector. Remarkably, supersensitivity can be observed even with inefficient but heralded single photon sources.

As the number of modes increases, we expect that a single NRD will be insufficient to capture the required information that allows the device to be supersensitive.  A future analysis of the scaling of the number of NRDs necessary to maintain supersensitivity would be useful to determine if this device would then imply a truly scalable quantum measurement device that exceeds the precision of classical sensors.

\ack
C.Y. would like to acknowledge support 
from an Economic Development Assistantship from the the Louisiana State University System Board of Regents. S.A. and J.P.D. would like to acknowledge support from the  ARO, AFOSR, DARPA, NSF and NGAS. T.B. is supported by the Shanghai Research Challenge Fund; New York University Global Seed Grants for Collaborative Research; National Natural Science Foundation of China (Grant No. 61571301); the Thousand Talents Program for Distinguished Young Scholars (Grant No. D1210036A); and the NSFC Research Fund for International Young Scientists (Grant No. 11650110425); NYU-ECNU Institute of Physics at NYU Shanghai; and the Science and Technology Commission of Shanghai Municipality (Grant No. 17ZR1443600). J.P.O. acknowledges support from the Vannevar Bush Faculty Fellowship program sponsored by the Basic Research Office of the ASD(R\&E) and funded by the ONR through grant N00014-16-1-2008.

\appendix
\section*{Appendix A}
\setcounter{section}{1}

Here we calculate the entries of the QFI matrix $\mathcal{F}_{\vec{\boldsymbol{\varphi}}}$.  In general, they are specified by,

\begin{equation}
	[\mathcal{F}_{\vec{\boldsymbol{\varphi}}}]_{l,n}=4\left< \hat b_l^\dag {\hat b_l}\hat b_n^\dag {\hat b_n} \right> -4 \left< \hat b_l^\dag {\hat b_l} \right>\left< \hat b_n^\dag {\hat b_n} \right>.
\end{equation}
Computing the latter term first,
\begin{eqnarray}
\langle\psi| \hat b_{j}^{\dagger} \hat b_{j} |\psi\rangle
&=\sum_{q,l=1}^m V_{j,q}\bar V_{j,l}  \langle k |^{\otimes m}  \hat a_{q}^{\dagger} \hat a_{l} |k\rangle^{\otimes m}\nonumber \\
&=\sum_{q=1}^m | V_{j,q}|^2  \langle k |^{\otimes m}   \hat a_{q}^{\dagger} \hat a_{q} |k\rangle^{\otimes m} \label{eq:onemode} \\
&= \sum_{q=1}^m \frac{1}{m}\cdot k \nonumber \\
&= k. \nonumber
\end{eqnarray}
Hence,
\begin{equation}
4 \left< \hat b_l^\dag {\hat b_l} \right>\left< \hat b_n^\dag {\hat b_n} \right>=4k^2.
\end{equation}
Meanwhile,
\begin{eqnarray}
\langle\psi| \hat b_{l}^{\dagger} &\hat b_{l} \hat b_{n}^{\dagger} \hat b_{n}|\psi\rangle \\
&=\sum^m_{i,j,q,p=1} V_{l,i} \bar V_{l,j}V_{n,q}\bar V_{n,p}  \langle k |^{\otimes m} \hat a_{i}^{\dagger} \hat a_{j}  \hat a_{q}^{\dagger} \hat a_{p} |k\rangle^{\otimes m} \\
& =\sum^m_{q,p, q\neq p} V_{l,p} \bar V_{l,q}V_{n,q}\bar V_{n,p}  \langle k |^{\otimes m} \hat a_{p}^{\dagger} \hat a_{q}  \hat a_{q}^{\dagger} \hat a_{p} |k\rangle^{\otimes m} \nonumber \\
& \hspace{20pt} + \sum^m_{q,p} |V_{l,p}|^2| V_{n,q}|^2  \langle k |^{\otimes m} \hat a_{p}^{\dagger} \hat a_{p}  \hat a_{q}^{\dagger} \hat a_{q} |k\rangle^{\otimes m}.
\end{eqnarray}
The second term here is essentially just the square of Eq.~(\ref{eq:onemode}), and equal to $k^2$, hence,
\begin{eqnarray}
& = k^2+\sum^m_{q,p,q \neq p}( V_{l,p} \bar V_{l,q}V_{n,q}\bar V_{n,p}) k(k+1)
\end{eqnarray}
Then we have to evaluate,
\begin{equation}
	\sum_{q,p=1,q \neq p}^{m}( V_{l,p} \bar V_{l,q}V_{n,q}\bar V_{n,p}).
\end{equation}
Rewriting $\omega=e^{2\pi i/m}$ as the first $m$th root of unity,
\begin{eqnarray}
&=\frac{1}{m^2}\sum\limits_{q,p=1,q \neq p}^m {{\omega ^{(l - 1)(p - 1)}}{\omega ^{ - (l - 1)(q - 1)}}{\omega ^{(n - 1)(q - 1)}}{\omega ^{ - (n - 1)(p - 1)}}}\\
&=\frac{1}{m^2}\sum _{ q,p=1,q \neq p}^{ m }{ { { [{ \omega ^{ (p-q) } }] }^{ (l-1) } }{ { [{ \omega ^{ (q-p) } }] }^{ (n-1) } } } \\
&=\frac{1}{m^2}\sum _{ q,p=1,q \neq p}^{ m }{ { { [{ \omega ^{ (p-q) } }] }^{ (l-1) } }{ { [{ \omega ^{ -(p-q) } }] }^{ (n-1) } } } \\
&=\frac{1}{m^2}\sum _{q,p=1,q \neq p}^{ m }{ { { [{ \omega ^{ (p-q) } }] }^{ l-n } } }
\end{eqnarray}
If $l-n=0$, i.e. for the diagonal entries of $\mathcal{F}^{\rm{quant}}_{\boldsymbol{\varphi}}$, the summand is 1 and hence the sum evaluates to $m^2-m$.  For the off-diagonal terms, let $l-n=k$, so
\begin{eqnarray}
=\frac { 1 }{ { { m^{ 2 } } } } \sum _{ q,p=1,q \neq p}^{ m }{ { { [{ \omega ^{ (p-q) } }] }^{ k } } }\\
=\frac { 1 }{ { { m^{ 2 } } } } \sum _{ q,p=1,q \neq p}^{ m }{ { { [{ \omega ^{ k } }] }^{ (p-q) } } }
\end{eqnarray}
Let $p-q=r$, and note that $r\neq 0$.   There are $m$-many $\{p,q\}$ pairs whose difference is $r$ (or congruent to $r$(mod $m$), since $\omega$ is a $m$th root of unity).  Thus the sum reduces to,
\begin{eqnarray}
=\frac { 1 }{ { { m^{ 2 } } } } [m\sum _{ r=1 }^{ m-1 }{ { { [{ \omega ^{ k } }] }^{ r } }] }\\
=\frac { 1 }{ { { m^{ 2 } } } } [m[-1]]\\
=-\frac { 1 }{ { { m } } },
\end{eqnarray}
where we have used the fact that the sum over all powers of any $k$th root of unity is equal to one. Thus, the terms of the QFI matrix simplify to,
\begin{equation}
	[\mathcal{F}_{\vec{\boldsymbol{\varphi}}}]_{l,n}=
    \begin{array}{lc}
  \Bigg\{ & 
    \begin{array}{lc}
      4k(k+1)\cdot\frac{m-1}{m} & l=m \\ 
      4k(k+1)\cdot-\frac{1}{m} & l\neq m
    \end{array}
\end{array}.
\end{equation}

\end{document}